\documentclass[runningheads,a4paper]{llncs}
\usepackage[pdftex]{graphicx}

\usepackage[caption=false,font=normalsize,labelfont=sf,textfont=sf]{subfig}
\usepackage{url}

\usepackage{float}
\usepackage{multirow}
\usepackage{array,booktabs,ragged2e}
\newcolumntype{L}[1]{>{\RaggedRight\arraybackslash}p{#1}}
\newcolumntype{R}[1]{>{\RaggedLeft\arraybackslash}p{#1}}

\usepackage{cite}
\usepackage{amsmath,amssymb,amsfonts}
\usepackage{algorithmic}
\usepackage{graphicx}
\usepackage{textcomp}
\usepackage{xcolor}
\def\BibTeX{{\rm B\kern-.05em{\sc i\kern-.025em b}\kern-.08em
		T\kern-.1667em\lower.7ex\hbox{E}\kern-.125emX}}
	
\begin{document}
	
	\title{Continuous Compliance using Calculated Event Log Layers
	}
		
	\author{Teemu Lehto\inst{1}\orcidID{0000-0002-1332-1853} \and Johan Myrberger\inst{2} \and Apoorva Pandey\inst{3} }
	\authorrunning{Teemu Lehto, Johan Myrberger, Apoorva Pandey}
	\institute{
		QPR Software Plc, Finland
		\and
		Ericsson, Sweden
		\and
		Ericsson, India
	}
	
	\toctitle{Continuous Compliance using Calculated Event Log Layers}
	
	\tocauthor{Teemu Lehto, Johan Myrberger, Apoorva Pandey}

	\maketitle

	\begin{abstract}
		Compliance has traditionally been a reactive activity, where directives and guidelines have been formally documented and, to a large extent, been assumed to be followed. This traditional approach does not always work, and failure to be compliant has various consequences for many companies, ranging from penalties and fines to lockout of business opportunities and markets. Continuous compliance using calculated event log layers brings compliance into a continuous mode, where risk vectors become visible in real-time data and actions are taken continuously and proficiently. Our model of process mining utilizes calculated events to implement this continuous form of compliance management. Continuous compliance refers to observing possible compliance violations and also includes the ability to provide real-time and prompt feedback to the responsible Compliance officers who will take the preventive actions in parallel.
		Continuous compliance is also explored from the viewpoint of Trade Compliance with a case study.
		
	\end{abstract}
	

	\section{Introduction}
	
Processes in an organization have a dual role; to execute and create value efficiently and avoid compliance violations.
Process compliance, or process adherence, as a phrase is often used to assess if the intended process was followed or not, without looking closer at the objectives of following the process. Will not adhering to the process have a cost in terms of efficiency or in terms of external regulatory aspects? In this paper, we are interested in the latter, although the methodology presented is suitable in both cases.

Failure to be compliant with external requirements has various consequences for a company, ranging from penalties and fines to lockout of business opportunities and markets. Hence the problem of being compliant enough is a critical problem area for any company. 
Compliance to external aspects is mainly a reactive effort where directives and guidelines have been formally documented and largely assumed to be followed. The follow-up comes to a large extent from audits (internal and external). In a business environment that is fairly stable and with long cycles, this has been the main compliance model. Meanwhile, e.g., in software production flows (“DevOps”), which have much shorter clock cycles, the notion of “continuous compliance” is used. When the end-to-end business processes and value flows of a company get more agile and meet an ever increasingly competitive environment, the current “slow” compliance approach becomes a burden and a risk.

	\begin{figure}[htbp]
		\centerline{\includegraphics[width=0.9\textwidth]{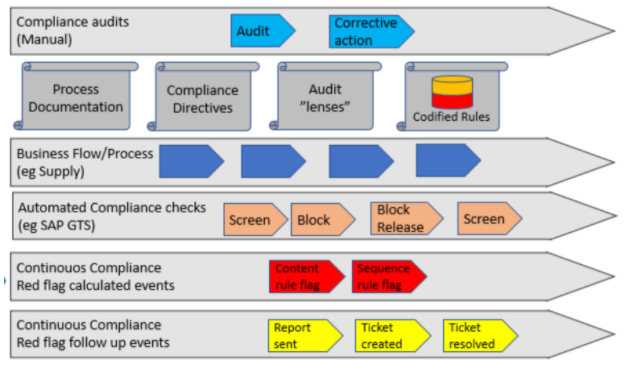}}
		\caption{Compliance approach augmented with the calculated event logs}
		\label{PIC_EventLogLayers}
	\end{figure}

	In this paper, we explore the application of continuous compliance and auditing in the general context and specifically in a trade compliance context. We divide the compliance problem statement into compliance during design time, compliance during run time, and compliance using audits (post-process execution). These phases are identified in the model outlined in the Fig.~\ref{PIC_EventLogLayers}, which shows our compliance approach augmented with the calculated event logs.
	
	The design-time aspects of compliance are captured in process documentation when designing the process, which is supposed to take into account the regulations and other compliance directives the organization needs to comply with. These are commonly based on various regulatory, legal, and other compliance frameworks such as ISO certifications together with other internal directives and steering documents collected into the corporate management system as the compliance directives. These compliance directives are taken into account in the design-time phase of processes when the intended process is designed and documented in process documentation.
	From this viewpoint, we in this paper bring forward a number of enhancements to solve the compliance business problem using process mining methodologies. We also identify a number of transformations in the business flow design that our approach opens up the need for.
	
	We problematize the current, classical compliance approach. It is clear from the Fig.~\ref{PIC_EventLogLayers} that the classical approach follows the sequential flow. Though sequential processing has steadily kept increasing, one can not ignore an impenetrable wall that is fast approaching.
	Multi or parallel processing is fast becoming the standard for performance optimization.Sequential processing does not entail the intricate job of managing data flow and distribution of work to various processes, which adds redundant lead time.
	
	This paper is organized as follows: Section \ref{Related Work} presents previous work in the area. Section \ref{Methodology} presents our methodology for doing Continuous Compliance, followed by an example case study in section \ref{case}. Further discussion and summary is presented in the last Section \ref{Summary}.

	\section{Related Work}
	\label{Related Work}
	
	The interpretation of the external compliance requirements and regulations are an integral part of the Management Control System (MCS) of a company \cite{MARGINSON}.

The audit phase is often reactive and “post-mortem.” At certain intervals an external or internal audit is performed, and the findings are dealt with using corrective actions. The methodology of the audit, including what signals and symptoms the auditor looks at, are part of the auditing system. 

Corporate processes and value flows are executed with IT tools support. When the processes are executed, they create data and digital footprints that can be used for process mining purposes to identify improvement opportunities. The central information objects used in process mining are the events, which is the trail that a case that flows through the process leaves behind. Process mining as a discipline contains many examples of how process efficiency can be increased, using process mining as a methodology of analysis \cite{MANIFESTO}. Another use case for process mining is to solve the company problem of compliance, where process mining has been used to monitor processes continuously. An overview of conducting compliance monitoring in business processes is presented in \cite{LY}.
In our paper, we present a method that touches the organizational aspects to be taken care of when setting up a continuous compliance system.
A methodology and tool support for managing business rules in organizations has been presented \cite{BAJEC}, however, that paper does not show the idea of using process mining. 

A framework for anomaly detection has been presented in \cite{BEZERRA}. However, our method shows how the detected anomalies should be presented as an additional layer together with the anomaly follow-up activities to give users better understanding of compliance activities. We use a Compliance Request Language to formally specify the compliance rules \cite{ELGAMMAL} and store the rules into a compliance rule registry to enable runtime detection of process violations as presented in \cite{AWAD}.

The solution to the compliance problem can be sought using two different approaches; to design an intended execution path that is by design safe enough, or to educate and support the executioner of the nature of the regulatory risks and delegate the risk mitigation beyond the process design. It can be noted that the earlier version (design an intended execution path) seems suitable for known, repetitive tasks, while a more delegated approach seems suitable for less repetitive and hence more unknown situations.

Furthermore, the company is not only faced with avoiding compliance risks, but also with the task to document the risk mitigation and compliance activities implemented, and to withstand audits, often by external auditors. Risk mitigation and compliance activities include the discovery of the root causes for irregularities that are found repeatedly. A process mining model built using the multi-layered methodology presented in our paper can be further analyzed with automated root cause analysis algorithms such as described in \cite{LEHTO} to support process improvement.

	\section{Continuous Compliance Methodology}
	\label{Methodology}

	Our method for continuous compliance using calculated event log layers consists of the following steps. 
		
	\subsection{Select the processes}

Select one or more business processes that will be the subject of the continuous compliance activity. It is mandatory to identify a business process in order to use process mining techniques for compliance monitoring. Trade compliance rules are typically related to multiple business processes with case ids such as sales order header id, sales order line id, shipment header id, shipment line id, purchase order header id, purchase order line id, good issue id, and goods receipt id. Each compliance rule specified in the previous step is mapped to the corresponding process.

	\subsection{Specify Compliance rules}
	
	It is important to find the intersection between a business process and a set of compliance directives. We suggest using a Compliance Request Language such as \cite{ELGAMMAL} to formally specify the compliance rules. Some of the compliance violations can be detected already at the very beginning of a process. For example, content rules are used during the trade compliance checks to detect business partners that are listed in the US sanctions list. Some violations occur or can be detected only after the process has started, such as a shipment being sent before a sales order validation has taken place.
	
	\subsection{Identify and specify relevant compliance rules and anti-patterns}
	
	Relevant compliance rules are found in the intersection of the selected business processes and targeted compliance areas, such as trade compliance.
The compliance rules and anti-patterns can be related to the sequence of certain business process events (“event a shall happen before event b,” or “event c shall not occur before event b”) or to specific data elements used in the process.

\subsection{Identify other compliance checks}

In certain intersections of business processes and compliance processes, there might already be an active compliance support implemented, beyond the pure audits. For example, a global trade services system such as SAP Global Trade Services (SAP GTS), may be in use to ensure compliance of a supply business flow in large organizations. 

\subsection{Extract data from ERP and create process mining model}

We start by identifying the different data set in the ERP system, such as Master Data(SAP ONE) and Trade compliance data(SAP GTS system). Once we have information from source systems, we then connect to the database and read them from different landing layers/schema, and continue to join them together to form the unified data set. Once this is achieved, we use the classic approach to form Case attribute and Event tables to form the process mining model.
In our methodology, we have used Structured Query Language (SQL) for this.
As a result, we have the business process flow layer.

\subsection{Extract Compliance Checks}

Basic compliance checks are done in Compliance ERP, such as the SAP GTS system. These events will form the Compliance check layer. 

\subsection{Runtime Compliance Rule Monitoring}

One key part of the suggested and implemented methodology is to generate an additional set of events, an event log, out of the governance rules and patterns.
We take an identified business process (eg sourcing), and we reuse existing structural knowledge around compliance rules (an activity that shall be done to be compliant) and anti-patterns (patterns that should not happen as they either are not compliant in themselves or are closely correlated to possible non-compliancies).
Evaluate the compliance rule performance for each case and add compliance rule violations with the exact timestamps of when the violation took place. Generate Compliance Rule violation layer.

\subsection{Follow-up actions}

Initiate actions based on the detected compliance violations. The sending of each violation is added to the Compliance Follow-up Layer. We suggest following
\begin{itemize}
	\item Send an email report to the responsible employees. This serves as the first automated report and helps the development team to start discussions about further mitigation actions and the validity of rules.
	\item Create a new ticket to an incident management system. This creates the initial formal need for the responsible persons to act and start investigations.
	\item Start a Robotic Process Automation (RPA) bot to help responsible persons in the investigations.
\end{itemize}

\subsection{Create Multi-Layer process models}

Generate a full process model based on all the above layers to see how the business operations, compliance checks, business rule violations, and compliance mitigation actions work in parallel to ensure continuous compliance. These models are used in two levels: individual compliance anomaly level and generic business process improvement level. On the individual compliance anomaly level, the benefits include:
\begin{itemize}
	\item Provide an up-to-date view into the individual case, with the augmented information of compliance checks, compliance rule violation time, follow-up actions, current situation, collaborative information from the follow-up actions.
	\item Employees from the compliance department can align their actions with the employees in the business units.
	\item Corrective actions can be taken to eliminate the risk before the formal audits.
	\item Individual cases can be blocked (stopped) to avoid further problems.
\end{itemize}
On the business process improvement level, the benefits include:
\begin{itemize}
	\item Provide dashboards for monitoring the compliance rule anomalies amounts
	\item Process mining models to analyze the lead times between anomaly violation, anomaly detection, follow-up action start, and follow-up resolution.
	\item Find root causes for systematic anomalies and inefficiencies.
\end{itemize}

	\section{Case Study} 
	\label{case}
	
	In this section, we show how the proposed methodology is implemented in a scenario in the trade compliance context. We first present a running example followed by an actual industrial case presentation.

\subsection{Trade Compliance Running Example} 

Trade compliance is a subset of the overall compliance problem, the subset where trade regulations are monitored and ensured. The Table \ref{TABLE_BusinessEvents} shows an example of supply chain process events. In our scenario, the supply flow is executed and logged in an ERP system, where the events in the Table \ref{TABLE_BusinessEvents} are found. In this trade compliance case, the compliance is enhanced in run time using a supporting system, which scans the supply cases and checks for specific rules such as the destination country for the delivery. If certain criteria are fulfilled, a hard block is applied to the process execution in the ERP system, and the block needs to be removed before the process execution can continue. A number of sample events from the run time compliance system are shown in the Table \ref{TABLE_ComplianceScreeningEvents}

	\begin{table}[htbp]
		\caption{Running Example Data - Business Flow Events}
		\begin{center} \begin{tabular}{|c|c|c|c|} \hline
				\textbf{CaseId} & \textbf{Activity} & \textbf{TimeStamp} \\ \hline
				C01 & Sales order received & 2021-06-18 \\ \hline
				C01 & Delivery created & 2021-06-20  \\ \hline
				C01 & Shipment started & 2021-06-21  \\ \hline
				C01 & Completed & 2021-06-22  \\ \hline
				
				C02 & Sales order received & 2021-07-18  \\ \hline
				C02 & Shipment started & 2021-07-20  \\ \hline
				C02 & Delivery created & 2021-07-21  \\ \hline
				C02 & Shipment cancelled & 2021-07-22 \\ \hline
				C02 & Shipment created & 2021-07-23 \\ \hline
				C02 & Completed & 2021-07-24 \\ \hline
			
			\end{tabular}
			\label{TABLE_BusinessEvents}
	\end{center} \end{table}

	\begin{table}[htbp]
		\caption{Running Example Data - Compliance Screening Events}
		\begin{center} \begin{tabular}{|c|c|c|c|} \hline
				\textbf{CaseId} & \textbf{Activity} & \textbf{TimeStamp} \\ \hline
				C01 & Trade compliance screening & 2021-06-19 \\ \hline
				C02 & Trade compliance screening & 2021-07-19  \\ \hline
			\end{tabular}
			\label{TABLE_ComplianceScreeningEvents}
	\end{center} \end{table}
	
	A set of calculated irregularity events is generated according to our methodology. These events are generated based on a set of business rules, which can be a function of specific event sequence combinations or a function of specific case attributes.
In the scenario of trade compliance, two example event sequence rules are found in the Table \ref{TABLE_RuleRegistry}.

When the sequence rules are applied to the business flow events, we get the calculated compliance violation events listed in the Table \ref{TABLE_IrregulationEvents}. These calculated irregularity events are communicated in a report format, and a follow-up case for each regularity is generated in a ticketing system. These corrective actions are also stored as events shown in the Table \ref{TABLE_FollowupEvents}, and become part of the overall aggregated process mining model.

	\begin{table}[htbp]
		\caption{Running Example Data - Compliance Rule Registry}
		\begin{center} \begin{tabular}{|c|c|c|c|} \hline
				\textbf{RuleId} & \textbf{Rule expression}  \\ \hline
				R01 & 'Shipment started' can only occur after 'Delivery created' \\ \hline
			\end{tabular}
			\label{TABLE_RuleRegistry}
	\end{center} \end{table}

	\begin{table}[htbp]
		\caption{Running Example Data - Compliance Irregulation Events}
		\begin{center} \begin{tabular}{|c|c|c|c|} \hline
				\textbf{CaseId} & \textbf{Activity} & \textbf{TimeStamp} \\ \hline
				
				C02 & Continuous Audit finding; R01 violation & 2021-07-21 \\ \hline
				
			\end{tabular}
			\label{TABLE_IrregulationEvents}
	\end{center} \end{table}
	
	\begin{table}[htbp]
		\caption{Running Example Data - Compliance Follow-up}
		\begin{center} \begin{tabular}{|c|c|c|c|} \hline
				\textbf{CaseId} & \textbf{Activity} & \textbf{TimeStamp} \\ \hline
				
				C02 & Compliance report sent & 2021-07-21  \\ \hline
				C02 & Compliance Incident created & 2021-07-22  \\ \hline
				C02 & Compliance Incident resolved & 2021-07-22  \\ \hline

			\end{tabular}
			\label{TABLE_FollowupEvents}
	\end{center} \end{table}
	
\begin{figure}[htbp]
		\centerline{\includegraphics[width=0.9\textwidth]{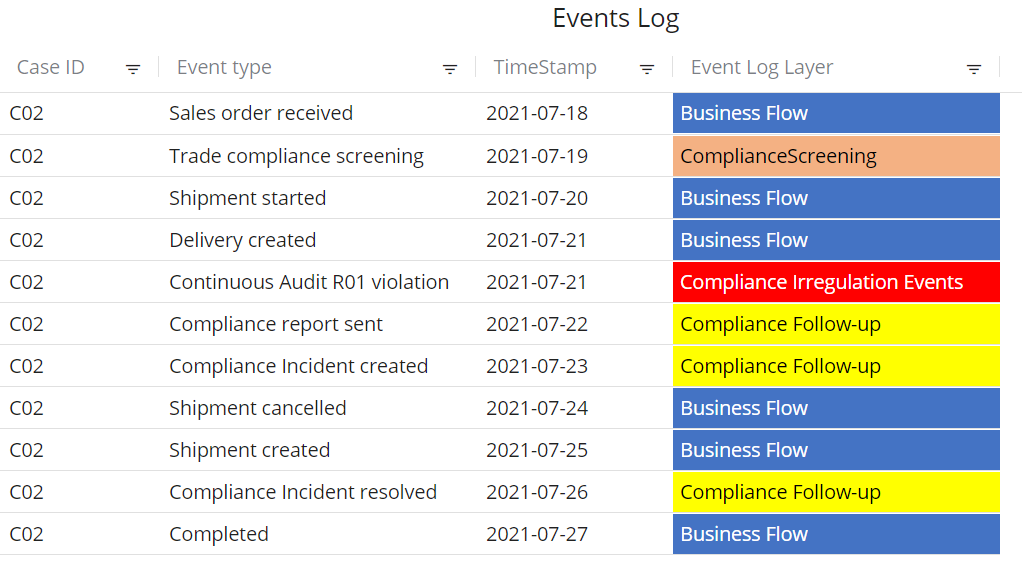}}
		
		\caption{Event Log with multiple Event Log Layers}
		\label{PIC_EventLog_Color}
	\end{figure}

	An example of showing these different event log layers with color codings is shown in the Fig.~\ref{PIC_EventLog_Color}. The process starts with the blue Business Flow event 'Sales order received', followed by a successful trade compliance check. The creation of shipment before the delivery violates a compliance sequence rule, resulting in compliance audit violation as well as the start of a compliance follow-up workflow. Follow-up actions lead to canceling of the shipment and creation of a new shipment. Having all these event log layers in the same process mining model helps a large organization to ensure follow-up action effectiveness.
	
	\subsection{Trade Compliance Industrial Case}
	
	Our methodology hes been implemented in an international organization to the supply process flow from a trade compliance viewpoint. Methodology combines a layer of calculated red flag events with other process mining events. We look at the rule constructs that define the calculated events and the benefits of using an event layer of calculated red flag events. This methodology enables a continuous audit approach, which complements run-time compliance.
	
The introduction of a continuous audit trail requires a supporting process of continuous corrective actions and follow-up of the findings from the continuous audit, and we identify this as a layer in the layered event structure. Further, to answer the question of who the responsible parties are and how they will be notified and act on these anomalies, we are including functionalities like notifications and connecting to a ticketing system.

A number of business isseus were identified in the run time compliance system setup, ranging from excessive volume of compliance screenings for certain case types (resulting in system performance degradation risks) to the possibility to carry out activities in the ERP system despite that a block has been applied.

A weekly report of calculated red flag events has been established, providing the trade compliance organization with a new approach to proactively mitigate trade compliance risks. The set of calculated red flag events are further analyzed using process mining methodology and flowcharts, both as a separate set of events and as an aggregated set of events, stacked with the other Event Log layers as outlined in the Fig.~\ref{PIC_EventLogLayers}

\begin{figure}[htbp]
		\centerline{\includegraphics[width=0.8\textwidth]{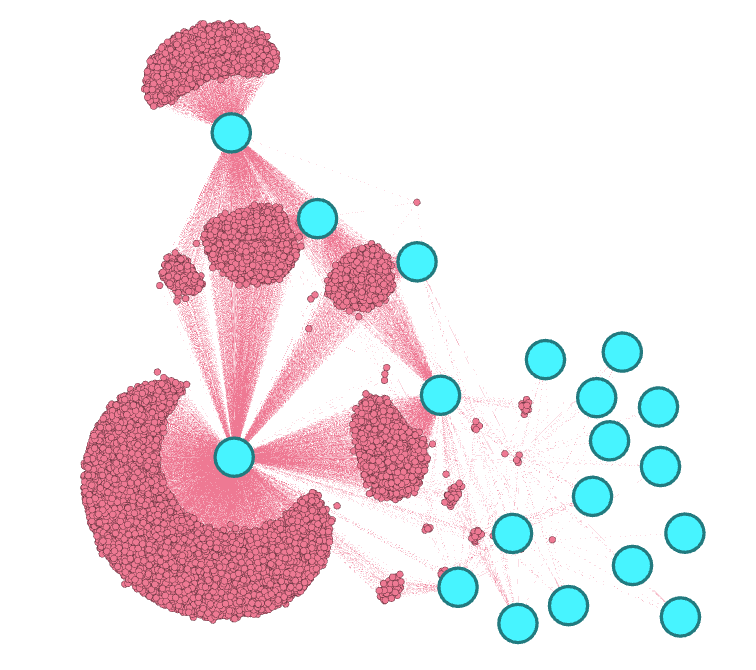}}
		
		\caption{Compliance Rule Violation Network.}
		\label{PIC_ComplianceRuleViolationNetwork}
	\end{figure}

The red flag events and the cases that contain these events are presented in a graph format with two types of nodes in the Fig.~\ref{PIC_ComplianceRuleViolationNetwork}. The red flag irregularity events (large, blue, nodes) and the individual supply flow cases (small, red, nodes), By applying a force atlas graph layout algorithm, the supply flow cases form clusters that relate to one or more of the continuous audit rules. The visually observable clusters of cases suggest a hypothesis for further analysis, such as the larger clusters of case nodes might be due to systemic problems in the intended process, and small clusters of case nodes indicate more intermittent irregularity combinations.
	
	Process mining represents the missing link between the analysis of big data and business process management. With the calculated and layered event model, we can implement a continuous compliance and continuous audit approach to detect the irregularities, send a notification to the accountable users and connect to the ticketing tool to log incidents so the corrective actions can take place seamlessly and structured. A person working as the Head of Trade Compliance Operational Development Director said: \textit{"Instead of having the audit every year or each month, now we conduct a full audit every week."}

	\begin{figure}[htbp]
		\centerline{\includegraphics[width=0.9\textwidth]{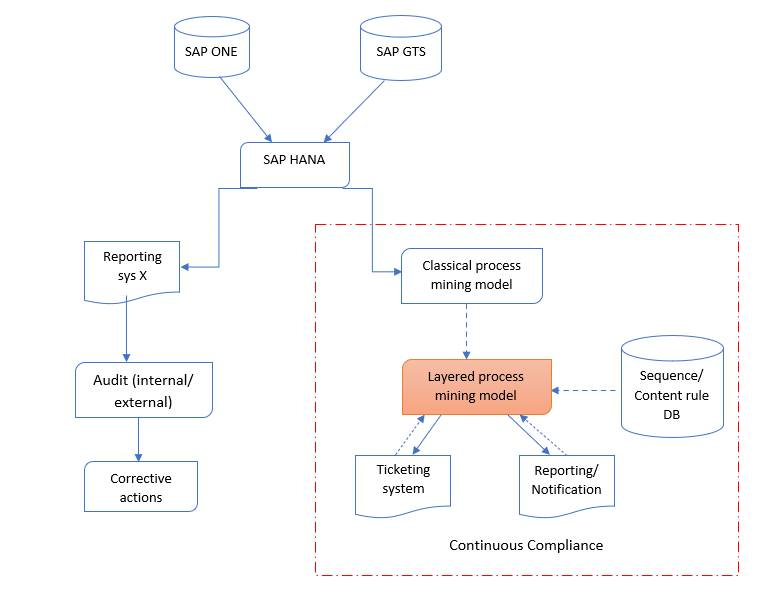}}
		\caption{Compliance Approach.}
		\label{PIC_ComplianceApproach}
	\end{figure}
	
	Fig.~\ref{PIC_ComplianceApproach} shows the system setup for the implemented continuous compliance approach connecting the ticketing system, rule registry, and reporting notifications to the layered process mining model. The business flows have a digitalized/automated aspect of run-time compliance. One used system is the SAP GTS (Global Trade Services), which has a set of features that integrate with the business process support in the ERP (like SAP). The GTS system executes scans using a set of defined rules like “in which country is the destination customer located,” and if the scan result meets certain criteria, an automatic block is applied, which prevents any further process execution of the business process in the ERP system, until the block is released.
	
	\section{Summary and Conclusions}
	\label{Summary}
	
In this paper, we have presented a methodology to establish continuous compliance operations using calculated event log layers. The method was presented in chapter 3, followed by a running example and a case study.

A layered event flow approach adds benefits to the classic compliance solution by aggregating the business process events together with the run-time compliance events. This in itself does not introduce any new process or event components, but the co-layered approach introduces some advantages.
The process execution activity and the run time compliance checks are usually the responsibility of two different sub-organizations. They might be synchronized on a management and resource allocation level, but the implications of the actual implementation and execution across the two areas of responsibility are not always obvious. By combining the two event flows into one process mining model and visualization, a joint reality is established, and glitches can jointly be worked out. Examples of these glitches include “process steps that are not allowed after a block has been set, and prior to the release of the block, are still executed ”or “screenings taking place more often than intended”.

In this paper, we have introduced calculated red flag events as a separate layer of event flows. This information is used to both provide notifications to compliance actors as well as utilizing these events in a process mining context, both as a stand-alone graph and in a multi-layered event aggregation model. We refer to this as continuous auditing.	Furthermore, we have noted that not all the event types in the process execution flow are relevant for the continuous audit approach. It is possible to use a simplified process mining model, with fewer event types, for working out the calculated events. This will reduce the computational and processing effort, which enables analysis of the larger business scope and/or analysis of a longer time span with the same computing resources.
To model the red flags as events in a process mining model has several advantages. It unlocks the usage of process mining methodology to analyze the event flow of red flags as such. Also, the ability to include the occurrence of red flag events as an aggregated part of the actual main process flow makes it easier to visualize and analyze the correlation. The identification and creation of these calculated events are done using a formalized set of red flag rules or codified continuous auditing rules. The codified auditing rules, used in the continuous auditing flow, become central in the overall corporate management system, and hence the codified rules must be governed and managed as a steering document. The introduction of continuous auditing, with continuous findings of irregularities in the process execution, leads to the need to also take continuous corrective actions. We introduce this as a separate process flow, adding to the layered events.

As presented in the case study chapter, we have used the methodology in a real business setting. Based on our experiences, the usage of our methodology leads to better compliance to rules with smaller costs, compared to traditional ad-hoc or unsystematic approaches. Results have been good, and the trade compliance officers see the benefit of this more systematic approach. The possibility to see continuous auditing activities and incident follow-up activities as separate event log layers together with the business flow helps the whole organization to quickly deal with any existing irregularity as well as to continuously develop the whole continuous compliance process.

	\subsubsection{Acknowledgements.}
	\label{acknowledgements}
	
	We thank Ericsson and QPR Software Plc for funding this work. The methodology presented in this paper has been implemented in a commercial process
	mining tool QPR ProcessAnalyzer.

	\vspace{12pt}
\end{document}